\documentclass[aps,prl,reprint]{revtex4-1}

\usepackage{graphicx,xcolor,soul}

\newcommand{\Schw}{Schwarzschild}

\newcommand{\beq}{\begin{equation}}
\newcommand{\eeq}{\end{equation}}
\newcommand{\bea}{\begin{eqnarray}}
\newcommand{\eea}{\end{eqnarray}}

\def\EE{{\cal E}}
\def\LL{{\cal L}}
\def\BB{{\cal B}}

\def\RS{\Sigma}
\def\DD{\Delta}

\providecommand{\dif}{\mathrm{d}} \def\d{\dif}

\begin{document}

\title{Radiative Penrose process: Energy Gain by a Single Radiating Charged Particle\\ in the Ergosphere of Rotating Black Hole}

\def\Opava{Research Centre for Theoretical Physics and Astrophysics, Institute of Physics, Silesian University in Opava, Bezru{\v c}ovo n{\'a}m.13, CZ-74601 Opava, Czech Republic}

\author{Martin Kolo\v{s}}
\affiliation{\Opava}
\email{martin.kolos@physics.slu.cz}

\author{Arman Tursunov}
\affiliation{\Opava}
\email{arman.tursunov@physics.slu.cz}

\author{Zden\v{e}k Stuchl{\'i}k}
\affiliation{\Opava}
\email{zdenek.stuchlik@physics.slu.cz}

\begin{abstract}
We demonstrate an extraordinary effect of energy gain by a single radiating charged particle inside the ergosphere of a Kerr black hole in presence of magnetic field. We solve numerically the covariant form of the Lorentz-Dirac equation reduced from the DeWitt-Brehme equation and analyze energy evolution of the radiating charged particle inside the ergosphere, where the energy of emitted radiation can be negative with respect to a distant observer in dependence on the relative orientation of the magnetic field, black hole spin and the direction of the charged particle motion. Consequently, the charged particle can leave the ergosphere with energy greater than initial in expense of black hole's rotational energy. In contrast to the original Penrose process and its various modification, the new process does not require the interactions (collisions or decay) with other particles and consequent restrictions on the relative velocities between fragments. We show that such a Radiative Penrose effect is potentially observable and discuss its possible relevance in formation of relativistic jets and in similar high-energy astrophysical settings.
\end{abstract}

\maketitle

\section{Introduction} \label{sec:intro}

The fundamental role of combined strong gravitational and magnetic fields for processes around BH has been proposed in \cite{Ruf-Wil:PRD:1975:,Bla-Zna:1977:MNRAS:}. There are convincing observational evidences that magnetic fields must be present in the vicinity of black holes (BHs) \cite{Daly:APJ:2019:}. Magnitude of magnetic fields around BHs may vary from few Gauss up to $10^8$~G, depending on the source generating the field. For stellar-mass BHs observed in X-ray binaries the characteristic magnitude goes up to $10^8$~G, while for supermassive BHs it is of the order of $10^4$~G. Since the energy densities of such magnetic fields are not sufficient to create significant contribution to the spacetime geometry, in realistic astrophysical situations the spacetime around a BH can be fully described by the Kerr solution of the Einstein field equations. Matter surrounding BH is usually assumed to be neutral, so that the rotating Kerr BH surrounded by neutral matter in an axially-symmetric configuration is considered as a standard model for compact sources as quasars or microquasars, but near the innermost circular orbit electric charges of particles constituting the globally neutral plasma start to be relevant. Due to the large value of the charge-to-mass ratio for elementary particles, near the BH horizon the electromagnetic Lorentz force can easily compete with BH gravity \cite{Stu-etal:2020:Universe:}.

Radiation reaction effects on charged particle dynamics have been studied in \cite{Tur-etal:2018:APJ:} for \Schw{} BH. Here we point out a fundamentally new effect that occurs due to the BH rotation. We show that in the BH ergosphere, a single charged particle can increase its energy due to its electromagnetic radiation. We call this new phenomenon the radiative Penrose effect, as it is caused by the extraction of BH's rotational energy due to capture of photons radiated with negative energy as related to distant observer. 

\section{Kerr spacetime and external magnetic field}

The Kerr metric in the Boyer--Lindquist coordinates and geometric units ($G=1=c$)
reads
\bea
 \d s^2 &=& - \left( 1- \frac{2Mr}{\RS^2} \right) \d t^2 - \frac{4Mra \sin^2\theta}{\RS^2} \, \d t \d \phi \nonumber\\
 && + \left( r^2 +a^2 + \frac{2Mra^2}{\RS^2} \sin^2\theta \right) \sin^2\theta \, \d \phi^2 \nonumber \\
 && + \frac{\RS^2}{\DD} \, \d r^2 + \RS^2\, \d\theta^2, 
 \label{KerrMetric} 
\eea
where
\beq
\RS^2 = r^2 + a^2 \cos^2\theta, \quad \DD = r^2 - 2Mr + a^2, \label{RSaDD}
\eeq
$M$ is the gravitational mass and $a$ denotes the spin, which fulfills the  condition $a \leq M$ for BHs. The event horizon of the BH is located at the surface $ r_{\rm H} = M + \sqrt{M^2 - a^2} $. The ergosphere is bounded by the surfaces: 
\beq
 r_{\rm H} < r_{\rm ergo} < M + \sqrt{M^2 - a^2 \cos^2\theta }. \label{ergosphereEQ}
\eeq
In the ergosphere the energy of a test particle or photon, as related to infinity, can become negative, allowing extraction of the BH's rotational energy \cite{Bar-Pre-Teu:1972:APJ:}.

We assume that the BH is immersed into external asymptotically uniform magnetic field with the strength $B$ oriented along the BH spin axis \citep{Wald:1974:PHYSR4:}. The Wald solution governing this field can be written as
\bea
 A_t &=& \frac{B}{2} \left( g_{t\phi} + 2 a g_{tt} \right) - \frac{Q}{2} g_{tt},   \\
 A_\phi &=& \frac{B}{2} \left( g_{\phi\phi} + 2 a g_{t\phi} \right) - \frac{Q}{2} g_{t\phi}.  
\eea
Here, $Q$ denotes the electric charge of the BH that is not necessarily zero in realistic situations. The charge of BH arises due to frame-dragging effect of magnetic field lines leading to the non-zero potential difference between the event horizon and infinity. This causes a selective accretion into the BH until the electric potential in a local frame is neutralized, i.e., $A^t = 0$. At the final stage of the selective accretion the covariant components of $A_\mu$  have the following simple form 
\beq   
A_t = \frac{B}{2} g_{t\phi}, \quad A_{\phi} =  \frac{B}{2} g_{\phi\phi},
\label{VecPotShort}
\eeq 
so, the BH possesses the induced Wald charge $Q_W = 2 a M B$ \citep{Wald:1974:PHYSR4:}. Timescale of the selective accretion is extremely  short for astrophysical BHs, therefore, the plausible scenario is the one with the induced BH charge \cite{Zaj-etal:2018:MNRAS:}. Note, that induced charge (different from $Q_W$) arises in arbitrary axially symmetric magnetic field configuration different from uniformity \cite{Zaj-etal:2018:MNRAS:}. The induced BH charge has a crucial importance for the process of acceleration of charged particles in BH vicinity \cite{Tur-etal:2019:ApJ:}.

\section{Radiation reaction formalism}

The presence of magnetic field modifies the canonical four-momentum  of a charged test particles according to $P_\mu = m u_\mu + q A_\mu$, where $m$, $q$ and $u^\mu$ are mass, charge and four-velocity of a test particle. Time and azimuthal components of the four-momentum correspond to the energy $E$ and axial angular momentum $L$ of the charged particle
\bea
- E &=&  P_t = m u_t + q A_t, \label{KillingEnergy} \\
 L &=& P_\phi = m u_\phi + q A_\phi. \label{KillingAngMom}
\eea
Both $E$ and $L$ are constants of the motion, if the electromagnetic radiation of the charged particle can be neglected. In general case, when the radiation of the charged particle is taken into account both $E$ and $L$ change in time. 

Dynamics of charged particle undergoing radiation reaction force in curved spacetime is governed by the DeWitt-Brehme equation \cite{Poisson:2004:LRR:}
\bea
&& \frac{D u^\mu}{d \tau} = \frac{q}{m} F^{\mu}_{\,\,\,\nu} u^{\nu}
+ \frac{2 q^2}{3 m} \left( \frac{D^2 u^\mu}{d\tau^2} + u^\mu u_\nu \frac{D^2 u^\nu}{d\tau^2} \right) \nonumber \\
&& + \frac{q^2}{3 m} \left(R^{\mu}_{\,\,\,\lambda} u^{\lambda}
+ R^{\nu}_{\,\,\,\lambda} u_{\nu} u^{\lambda} u^{\mu} \right)
+ \frac{2 q^2}{m} ~f^{\mu \nu}_{\rm \, tail} \,\, u_\nu,
\label{eqmoDWBH}
\eea
where in the last term of Eq.(\ref{eqmoDWBH}), the tail integral reads
\beq
f^{\mu \nu}_{\rm \, tail}  =
\int_{-\infty}^{\tau-0^+}
D^{[\mu} G^{\nu]}_{ + \lambda'} \bigl(z(\tau),z(\tau')\bigr)
u^{\lambda'} \, d\tau' .
\eeq
Here $R^{\mu}_{\,\,\,\nu}$ is the Ricci tensor, $G^{\mu}_{ + \lambda'}$ is the retarded Green function, and the integration is taken along the worldline of the particle $z$, i.e., $u^\mu (\tau) = d z^\mu(\tau) /d \tau $.

The Ricci term is irrelevant, as it vanishes in the vacuum metrics, while the tail term can be neglected for elementary particles, as shown in \cite{Tur-etal:2018:APJ:} and references therein. The equation (\ref{eqmoDWBH}) also contains the Schott term -- the third order time derivative of coordinates, which leads to the appearance of pre-accelerating solutions in the absence of external forces. However, one can effectively reduce the order of the equation by substituting the third order terms by derivatives of the external force. This is identical to imposing the Landau-Lifshitz method in its covariant form \cite{Lan-Lif:1975:CTF:}. The resulting equation of motion reads
\bea \label{curradforce}
&& \frac{D u^\mu}{d \tau} = \frac{q}{m} F^{\mu}_{\,\,\,\nu} u^{\nu}  \\
&& + \,  \frac{2 q^2}{3 m} \left(F^{\alpha}_{\,\,\,\beta ; \mu} u^\beta u^\mu + \frac{q}{m} \left( F^{\alpha}_{\,\,\,\beta}
F^{\beta}_{\,\,\,\mu} +  F_{\mu\nu} F^{\nu}_{\,\,\,\sigma} u^\sigma u^\alpha \right) u^\mu \right), \nonumber
\eea
where semicolon denotes the covariant coordinate derivative. Eq. (\ref{curradforce}) is a habitual second order differential equation, which satisfies the principle of inertia and does not contain runaway solutions \cite{Lan-Lif:1975:CTF:}. In the magnetized Kerr BHs case Eq.~(\ref{curradforce}) takes very complex and long form, so we do not give it explicitly here.

Detailed analysis of the motion of charged particles in the case of Schwarzschild BH immersed into an asymptotically uniform magnetic field was presented in \cite{Tur-etal:2018:APJ:}, where the equations of motion are given in separated form and the results of their numerical integration are obtained in typical situations. Motion of charged particles around magnetized Kerr BHs is more complex in comparison with the magnetized \Schw{} BH \cite{Ali-Gal:1981:GRG:,Tur-etal:2018:APJ:}. Below we demonstrate an extraordinary effect of energy gain of the radiating particle in the ergosphere of magnetized Kerr BHs caused by emission of photons with negative energy as related to distant observers.

\begin{figure*}
\includegraphics[width=\hsize]{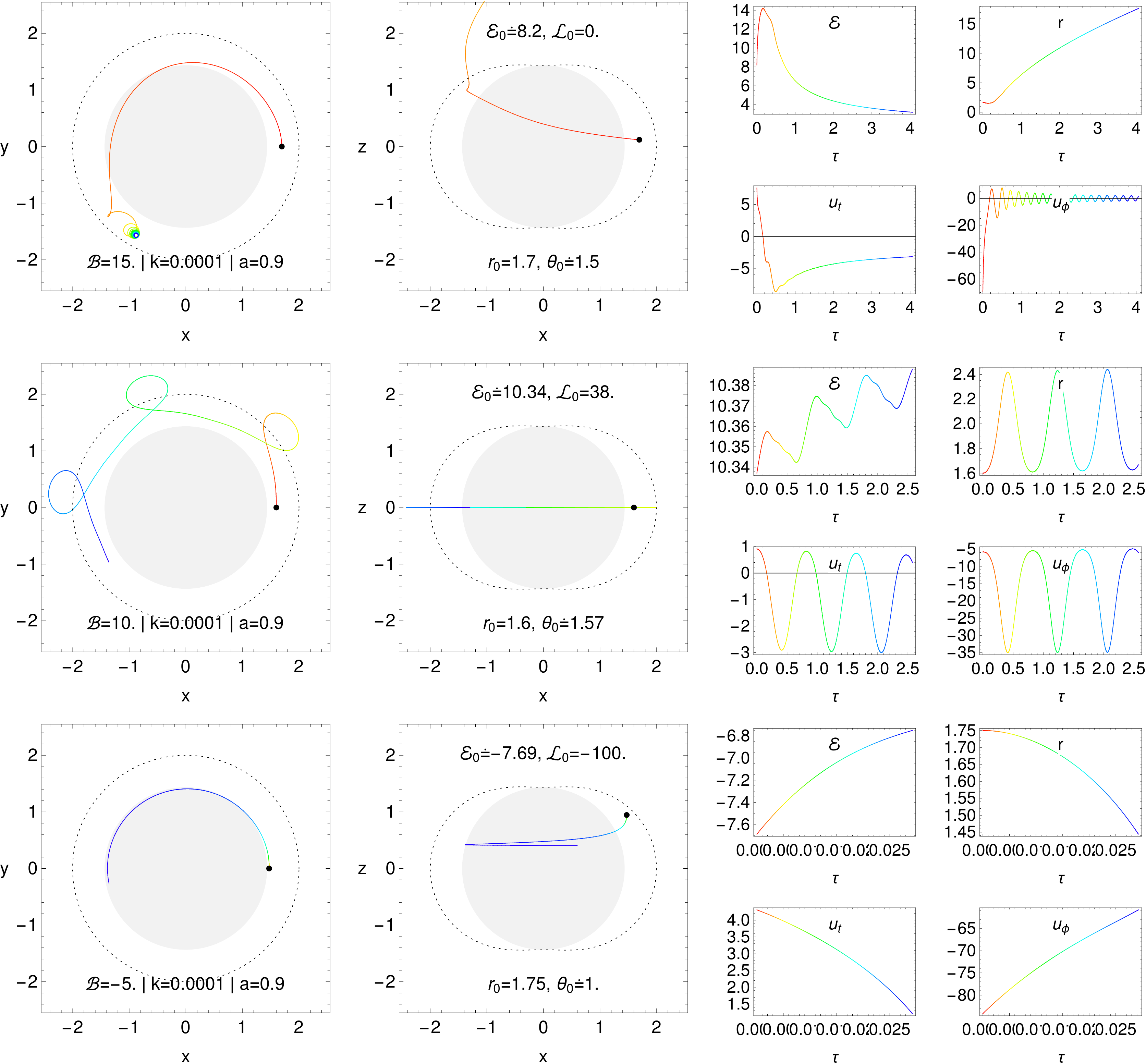}
\caption{Radiating charged particle trajectories demonstrating the radiative Penrose process for three different orbits. We show projections of particle trajectory into $xy, xz$ planes, particle specific energy $\EE$, radial coordinate $r$ and covariant component $u_t, u_\phi$, as functions of proper time $\tau$. Change of the colors denotes the evolution with the proper time, gray disk denotes BH interior, and dotted curve shows the boundaries of the egrosphere.
In the {\bf first} row the particle has been located little bit off equatorial plane and we see that radiative Penrose process occurs only at the beginning of particle's trajectory, when the particle is still in ergosphere and conditions for negative photon existence (\ref{negphoton}) are fulfilled. Particle is escaping ergoregion with increased energy, and during its escape from the BH along magnetic field line is loosing its energy by radiating photons with positive energy.
In the {\bf second} row we can see that negative photon existence, Eq.~\ref{negphoton}, is really the condition for radiative Penrose process: counter-rotating particle $u_\phi<0$ in equatorial plane is gaining energy inside ergosphere  ($u_t>0$, radiating photons with negative energy), while loosing energy when peek out of ergosphere ($u_t<0$, photons with positive energy).
In the {\bf third} row we can see counter-rotating particle energy gain before its capture by rotating BH.
\label{fig1}
}
\end{figure*}

\section{Negative energy photons\\ and charged particle radiation\\ inside the BH ergosphere}

Inside the ergosphere any particle (both massive and and massless) has to be co-rotating with the BH as seen by the asymptotic stationary observers ($u^\phi>0$). Such an observer can measure the energy of a particle negative, while the locally measured energy is always positive for a physical particle. A local observer, in this case, sees the particle in counter-rotating motion, if its $\phi$ component of covariant four-velocity is negative $u_\phi<0$. In the rotating BH ergosphere (\ref{ergosphereEQ}), radiated photons are attaining negative energies and negative angular momenta ($E_{\rm ph}<0, L_{\rm ph}<0$) related to distant observer when following conditions are met by the radiating particle
\beq
 u_t > 0, \quad u_\phi < 0. \label{negphoton}
\eeq
For details on the negative energy photons, see, e.g.\cite{Bar-Pre-Teu:1972:APJ:,Stu-Cha-Sch:2018:EPJC:}. The photons emitted by the relativistic charged particles in a synchrotron radiation can attain negative energy only if emitted backwards with respect to the BH rotation and radiating particle must be locally counter-rotating with $u_\phi<0$.

Photons with negative energy can never leave the ergosphere, being finally captured by the BH \cite{Stu-Cha-Sch:2018:EPJC:}. Emitted by the relativistic charged particles, these photons contribute to the spin down of the BH, which is equivalent to the extraction of the rotation energy of the BH. This is how the BH rotational energy can be extracted due to a single radiating charged particle. The photons emitted with negative energy are at the superradiance modes of the electromagnetic wave equations. However, we postpone the discussion of the detailed properties of the negative energy emission for further studies, concentrating instead on the dynamics of charged particles undergoing radiation reaction force by emitting negative energy synchrotron photons.  

In order to demonstrate the effect of the energy gain by a single radiating charged particle, one needs to integrate Eq.~(\ref{curradforce}) inside the ergosphere restricted by (\ref{ergosphereEQ}) for the Kerr metric (\ref{KerrMetric}). For simplicity we assume asymptotically uniform magnetic field, with the electromagnetic four-potential given by (\ref{VecPotShort}). For the convenience of the integration, we introduce the following independent parameters, characterizing the system:
\beq 
\EE = \frac{E}{m}, \quad \LL = \frac{L}{m}, \quad
\BB = \frac{q B M}{2 m}, \quad k = \frac{2}{3} \frac{q^2}{m M}, 
\eeq 
where $\BB$ is the magnetic parameter, which corresponds to the ratio of the Lorentz force to the gravitational force; parameters $\EE$ and $\LL$ are the specific energy and specific axial angular momentum of the charged particle and $k$ is the radiation parameter, characterizing the rate of the energy evolution. Results of the numerical integration of Eq.~(\ref{curradforce}) are demonstrated in Fig.~\ref{fig1} with initial conditions shown inside the plots. 

Radiative Penrose process allows the single radiating charged particle to escape from the ergosphere with energy exceeding its initial energy in expense of BH rotational energy. The relevance of the radiative Penrose process in dependence on magnetic parameter and BH spin is reflected in Tab.~\ref{tab1}. For particles following trajectories with curls, an oscillatory regime occurs where the energy gain periods are followed by energy loss. The energy gain process is also possible for charged particle trajectories finishing under the horizon.

\begin{table}[!ht]
\begin{center}
\begin{tabular}{l@{\quad} c@{\quad} l l l}
\hline
& & $E_0$ 	& $E_{\rm max}$ 	 & $E_{100}$ \\
\hline
mag. field $\BB$: 
& 14 & 7.7 & 13.2 & 2.8 \\
& 15 & 8.2 & 14.2 & 2.2 \\
& 16 & 8.7 & 15.2 & 2.5 \\
\hline
BH spin $a$: 
& 0.89 & 8.0 & 14.0 & 2.3 \\
& 0.90 & 8.2 & 14.2 & 2.2 \\
& 0.91 & 8.4 & 14.3 & 2.3 \\
\hline 
angular mom. $L_0$: 
& -10 & 7.1 & 14.4 & 2.3 \\
& 0 & 8.2 & 14.2 & 2.2 \\
& 10 & 9.4 & 14.0 & 2.3 \\ 
\hline
\end{tabular}
\caption{
Escaping orbit from Fig.~\ref{fig1} has been calculated using initial conditions $x^\mu=\{0,1.7,1.5,0\}, P_\mu\doteq\{8.2,0,0,0\}$ and parameters $a=0.9, \BB=15, k=10^{-4}$. We have particle initial specific energy $E_0$, maximal energy $E_{\rm max}$ increased by radiative Penrose process and energy at radius $r=100$ after the radiative loss $E_{100}$. 
In this table we changed one of the initial parameters for charged particle orbit, to demonstrate how radiative Penrose process depends on such parameter.
As it can be seen, the radiative Penrose process is especially relevant for strong magnetic fields and large BH spins.
\label{tab1}
} 
\end{center}
\end{table}

The amount of energy gain in this process depends not only on the parameters governing the electromagnetic forces $\BB$ and $k$, but also the length of stay of the charged particle in the ergosphere. It is different for trajectories with different initial conditions. If the charged particle starts its motion inside the ergosphere with relatively large pitch angle with respect to the equatorial plane (i.e. having non-vanishing $\theta$ component of the four-momentum), the trajectory lying inside the ergosphere will be short since the particle escapes in vertical direction along the magnetic field lines, the resulting energy gain will be relatively small, see first row of figures in Fig.~\ref{fig1}. If the trajectory lies very close to the equatorial plane with small or zero $p_\theta$, the particle moves in the ergosphere longer, which allows the particle to gain larger amount of energy due to radiation, see second row of figures in Fig.~\ref{fig1}. In general, the motion of a particle with ultra-relativistic energy is unstable inside the ergosphere, therefore, small change in $p_\theta$ (e.g. due to kick by other particle or photon), may force the particle to fall into BH or escape in the vertical direction along the magnetic field lines \citep{Tur-etal:2019:ApJ:,Stu-etal:2020:Universe:}.

Once the particle leaves the ergosphere, the energy of emitted photons with respect to distant observer turns to positive values and the charged particle continues its motion in the magnetic field loosing its energy due radiation reaction in the standard manner \citep{Tur-etal:2018:APJ:}. As one can see in the first row of Fig.~\ref{fig1}, the charged particle loses significant amount of its energy gained in the ergosphere after it leaves the ergosphere. In general, the rate of energy loss outside the ergosphere depends on the magnetic field strength and direction of the motion - see table in \citep{Tur-etal:2018:APJ:}. The radiation is maximal, if the direction of the motion is perpendicular to the magnetic field lines and minimal (basically zero) for the motion along the magnetic field lines (i.e. along the axis of BH rotation).


\section{Conclusions}

Extraction of the BH rotational energy due to radiating charged particle, in sharp contrast to the original Penrose process \cite{Penrose:1969:NCRS:} and its various modifications, does not require the assumption of interaction with other particles (collisions or decay). The original Penrose process, where the incident particle splits into fragments inside the ergosphere, requires the relative velocity between fragments to be relativistic ($>0.5$~c), which is the main argument against the realizability of the Penrose process in nature \citep{Bar-Pre-Teu:1972:APJ:}. In contrast, the Radiative Penrose process does not put such constraints. A single charged particle is able to gain energy inside the ergosphere and escape from the BH due to radiation-reaction force.

The characteristic timescale of synchrotron radiation by a single particle in magnetic field of strength $B$ can be estimated as \cite{Tur-etal:2018:APJ:}
\beq 
\tau_{\rm syn} \sim \left(1-\frac{2 G M}{r c^2} \right)^{-1} \frac{m^3 c^5}{q^4 B}.  
\eeq 
Due to cubic dependence on the particle mass, the radiation timescale is shorter for lighter particles and minimal for electrons (positrons), expected to produce the main part of the observed synchrotron emission from astrophysical sources. For SMBHs with typical magnetic field strength of $10^4$G, the characteristic synchrotron cooling timescale for electrons and positrons is of the order of a second (compare with the orbital timescale around SMBH $\sim 10^4$s). This is a small fraction of the orbital timescale, which implies that in realistic conditions the effect demonstrated on the numerical example in Fig.~\ref{fig1} is stronger by several orders of magnitude. 

As one can see, the effect can operate in a viable astrophysical conditions. Generally, the energy gain by radiative Penrose process implies decrease of expected radiation in the ergosphere, but increase of the radiating energy above the ergosphere. Total amount of extracted energy depends on the accretion rate of the BH, direction of the motion of the accretion disk and energy of synchrotron electrons or positrons. Described effect is potentially observable due to expected more powerful synchrotron emission on the edge of the ergosphere in contrast to the emission from surrounding matter. It would be useful to calculate the expected observational spectrum taking into account the synchrotron emission around the ergosphere and find the distinguishing marks characterizing the described mechanism. This effect could be observed as synchrotron peaks in SMBH (AGN, blazar, etc.) spectra, as well as in the spectra of microquasars, when high-resolution observational instruments become available as ATHENA, eXTP, etc.

Extraction of BH's rotational energy due to radiative Penrose process is potentially observable and can be relevant in many astrophysical BH systems.

\section{Acknowledgments}

The authors would like to acknowledge the Research Centre for Theoretical Physics and Astrophysics and Institute of Physics of Silesian University in Opava for institutional support.



\def\prc{Phys. Rev. C}
\def\pre{Phys. Rev. E}
\def\prd{Phys. Rev. D}
\def\jcap{Journal of Cosmology and Astroparticle Physics}
\def\apss{Astrophysics and Space Science}
\def\mnras{Monthly Notices of the Royal Astronomical Society}
\def\apj{The Astrophysical Journal}
\def\aap{Astronomy and Astrophysics}
\def\actaa{Acta Astronomica}
\def\pasj{Publications of the Astronomical Society of Japan}
\def\apjl{Astrophysical Journal Letters}
\def\pasa{Publications Astronomical Society of Australia}
\def\nat{Nature}
\def\physrep{Physics Reports}
\def\araa{Annual Review of Astronomy and Astrophysics}
\def\apjs{The Astrophysical Journal Supplement}
\def\aapr{The Astronomy and Astrophysics Review}
\def\procspie{Proceedings of the SPIE}



%

\end{document}